\documentclass[pre,aps,twocolumn,showpacs]{revtex4} 
\usepackage{epsfig} 
\newcommand{\Onecol} 
{\begin{widetext} 
\onecolumngrid} 
\newcommand{\Twocol} 
{\end{widetext} \twocolumngrid}
\begin{document}
\title{Coherence Resonance in an Autapse Neuron Model with Time Delay}
\author{$^1$Gautam C Sethia}
\email{gautam@ipr.res.in}
%\affiliation{Institute for Plasma Research, Bhat, Gandhinagar 382 428, India}
\author{$^2$J$\ddot{u}$rgen Kurths}
\email{jkurths@gmx.de}
\author{$^1$Abhijit Sen}
\email{abhijit@ipr.res.in}
\affiliation{$^1$Institute for Plasma Research, Bhat, Gandhinagar 382 428, India \\$^2$Institute of Physics, University of Potsdam, PF 601553, 14415 Potsdam, Germany}

\begin{abstract}
 We study the noise activated dynamics of a model {\it autapse} neuron system that consists of a subcritical Hopf oscillator with a time delayed nonlinear feedback. The coherence of the noise driven pulses of the neuron exhibits a novel double peaked structure as a function of the noise amplitude. The two peaks correspond to separate optimal noise levels for excitation of single spikes and multiple spikes (bursts) respectively. The relative magnitudes of these peaks are found to be a sensitive function of time delay. The physical significance of our results and its practical implications in various real life systems are discussed.
\end{abstract}
\maketitle
%\section{Introduction}

The constructive role of noise in the dynamics of complex systems is a subject of much current interest and activity. Important manifestations of such a behaviour are seen in basic phenomena like stochastic resonance (SR) \cite{gammaitoni98,wisenfeld98,wellens04}, coherence resonance(CR) \cite{pikovsky97} or noise-induced synchronization of dynamical systems \cite{zhou03,linder04}. Coherence Resonance (CR)  which refers to the resonant response of a dynamical system to pure noise, is closely related to the phenomenon of stochastic resonance and is sometimes also known as Autonomous Stochastic Resonance (ASR) \cite{andre97}. The effect, first noticed by Sigeti and Horsthemke \cite{sigeti89} in a general system at a saddle-node bifurcation, implies that a characteristic
correlation time of the noise-excited oscillations has a maximum for a certain noise amplitude. This has been clearly demonstrated for the classic FitzHugh-Nagumo neuron model\cite{pikovsky97} and shown to have a deep connection to the excitable nature of the system. CR can have important consequences for neurophysiology or other complex systems where a significant degree of order can arise through interaction with a noisy environment.  Past studies of CR have been mainly confined to simple systems whose noise-induced nonlinear outputs consist of impulsive excitations of a single kind e.g. spikes which have two characteristic time scales - a fast rise time and a longer decay time. This phenomenon has been found not only in various lab experiments, such as electronic circuits \cite{postnov99}, laser systems \cite{dubbeldam99, giacomelli00}, electrochemistry \cite{kiss03}, or BZ reactions \cite{miyakawa02} but also in natural systems such as ice ages in climatology \cite{pelletier03} or dynamos \cite{stefani05}.

\begin{figure}
\centerline{\includegraphics[width=8.0cm]{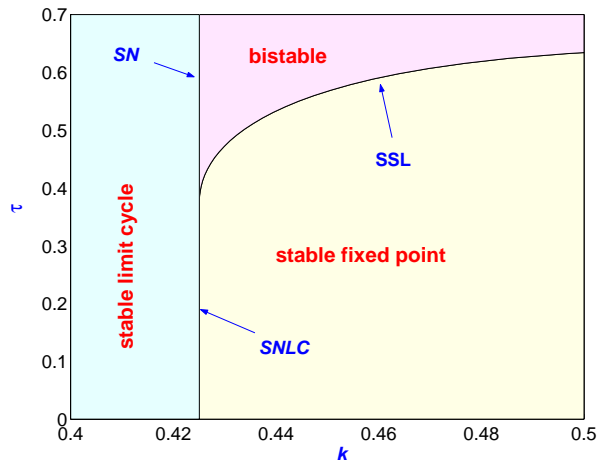}}

\caption{Stability diagram in the parameter space of $k$ and $\tau$. Note the various bifurcation boundaries demarcating the different stability regions. (SNLC: Saddle-Node on Limit Cycle bifurcation, SN: Saddle-Node bifurcation, SSL: Saddle-Separatrix Loop bifurcation)}
\label{bif}
\end{figure}

As is well known, excitable systems especially in neuroscience, can also have a more complex response in terms of various time scales, such as short time spikes and bursts (multi-spiking) with different temporal signatures\cite{izhikevich00, yacomotti99}. The nature of CR in the presence of different kinds of excitations (e.g. spikes and bursts) is not known and is one of the important objectives of the present study. In order to explore this question, we study the noise activated dynamics of a novel model neuron system that was recently presented in \cite{sethia06a}. The model consists of a subcritical Hopf oscillator with a time delayed nonlinear feedback. It contains the necessary ingredients to
reproduce most of the basic features of excitability that are displayed by standard models
such as the Hodgkin-Huxley system \cite{hodgkin52} and in addition provides a convenient means of studying the effect of time delay on neural dynamical behavior. 
%\section{Model Neuron}
The basic mathematical form of the model is,
\begin{equation}
\dot{z}(t)=\left[i(\omega+b\vert z(t)\vert ^{2})+\vert z(t)\vert ^{2}-\vert z(t)\vert ^{4}\right ]z(t)-kz^{2}(t-\tau) \label{feedback}
\end{equation} 
where $z=x+iy $ is a complex amplitude and the frequency of the oscillations is determined by $\omega$ and $b\vert z\vert^{2}$. The parameter $b$ which is called the shear parameter, determines how the frequency depends on the amplitude of the oscillations. $k$ is the magnitude of the feedback strength and $\tau$ is the time delay parameter. In the absence of the feedback term the oscillator is poised at the subcritical bifurcation point. The nonlinear feedback  term, which provides the basic excitable behavior, is modeled to mimic the structure of the recently discovered {\it autapse} neurons which have axons terminating back on their own dendrites \cite{tamas97,herrmann04}. The feedback is time delayed to account for finite propagation times of signals. The basic dynamical behaviour of the system is best 
captured in a two parameter bifurcation diagram (in $k$ and $\tau$ space) which was obtained in \cite{sethia06a} and is reproduced here as Fig. \ref{bif}. The diagram delineates the different bifurcation branches as well as the regions of stable limit cycle, stable fixed point and bistability. When subjected to an external periodic signal and noise the system displays both spike trains as well as multi-spiking (bursty) behaviour \cite{sethia06a}.
%,sethia06b}. 
For our present CR studies, we examine the temporal response of the neuron in the presence of only an external noisy stimulus, namely $f(t)=\sqrt{2D}\xi(t)$, where $\xi(t)$ is zero mean Gaussian white noise with intensity $D$. We confine ourselves to values of $k$ that are above the critical value of $k_c=0.42506$ and to $\tau$ values below the bi-stable region and choose different noise strengths $D$.\\

For each noise intensity $D$ we execute rather long simulations (100 datasets consisting of 200,000 time steps each with $\delta t=0.05$) and collect a large number of interspike intervals (ISI) $T$. Using this data we determine the standard parameter for coherence resonance $R$, which is given by 
the ratio of the mean of the interspike intervals ($\langle T \rangle$) and its standard deviation ($\sigma_{T}$)\cite{pikovsky97,mendez05,hu01,yacomotti99},
\begin{equation}
R=\langle T \rangle/\sigma_{T}
\end{equation} 
  
\noindent
If the data consists of a completely random (Poissonian) set of spikes then $R$ would have a value of unity, whereas a strictly periodic spiking train (maximum order) would make $R=\infty$. In general  $R > 1$ indicates a coherence, whereas values of $R$ that are significantly lower than unity indicate lack of coherence i.e. random dynamics. Our results for the present neuron model are shown in Fig. 2 where the upper block shows the CR and the lower block depicts the average interspike interval (ISI) as functions of the noise intensity $D$. The solid curves are for $\tau=0$ and the dashed curves correspond to finite time delay (in this case $\tau=0.3$). 
\begin{figure}
\centerline{\includegraphics[width=8.0cm]{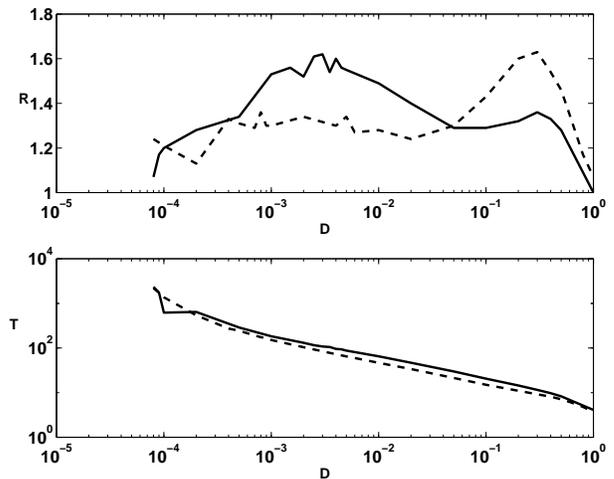}}
\caption{The CR measure $R$ and mean interspike interval $\langle T \rangle$ versus the noise intensity $D$. The solid curve is for $\tau=0$ and the dashed curve is for $\tau=0.30$. The feedback strength is $k=0.426$. The point is close to SNLC bifurcation branch. }
\label{r_d}
\end{figure}
We observe that unlike the standard CR results of a unimodal response curve \cite{pikovsky97} our present system produces a more complex response consisting of one well-expressed peak and, additionally, a broad shoulder in another region of noise intensities. The two-peaked structure is preserved in the presence of time delay with however one very substantial change - the relative amplitudes of the peaks are reversed. Without time delay there is  a broadband peak for small noise intensities $D$ around 0.003 and a shoulder with a minor peak at about $D=0.3$, whereas in the presence of time delay ($\tau=0.3$), we get a broad shoulder for small noise and  a rather narrowband peak at $D$ around 0.3. Note that in both cases the average ISI is decreasing with increasing $D$. Hence, the activity in the low noise region can be related to spiking (isolated spikes have a long interspike interval), whereas that in the strong noise region can be associated with bursting activity (several spikes within a burst leading to shorter ISIs). The bi-modal structure of the CR is a novel result which, to the best of our knowledge, has not been observed before and suggests that it can effectively distinguish between spikes and bursts and also provide a measure of their relative predominances. The presence of both spikes and bursts can be understood physically from the bifurcation diagram shown in Fig. 1. For $\tau=0$ and $k=0.426$ the system is located on the bottom horizontal line in the region of a stable fixed point. The introduction of noise pushes the system randomly into the stable limit cycle region from which it returns into the quiescent region along an extended trajectory on the unstable manifold. For low values of noise intensity the system barely makes it into the limit cycle regime and the resultant outcome is a single spike. At larger values of noise the system is driven deeper into the oscillatory region and spends more time executing several limit cycle circuits before returning to the quiescent state in a spiked manner. As a result the temporal trace now consists of an onset spike followed by several oscillations and a terminating spike - all of which taken together constitutes a burst. Such a phenomenon is well known in the literature as the circle/circle or parabolic bursting \cite{izhikevich00}. Spikes and bursts are optimally excited at two different values of the noise intensity as demonstrated by the CR curve. In the absence of time delay spikes appear to predominate as seen by the larger peak in the low noise regime and bursts have a relatively lower count of occurrences. \\

With the introduction of time delay the relative predominance of bursting appears to grow as a function of time delay. This can be seen from the dashed curves in the CR curve of Fig. 2 where there is a relative rise in the second peak at higher noise values. To get a better measure of this increase in bursting activity, we have also looked at the normalized probability distributions of the ISI for various values of $\tau$ at a fixed value of $D=0.03$ (Fig. 3). 
\begin{figure}
\centerline{\includegraphics[width=8.0cm]{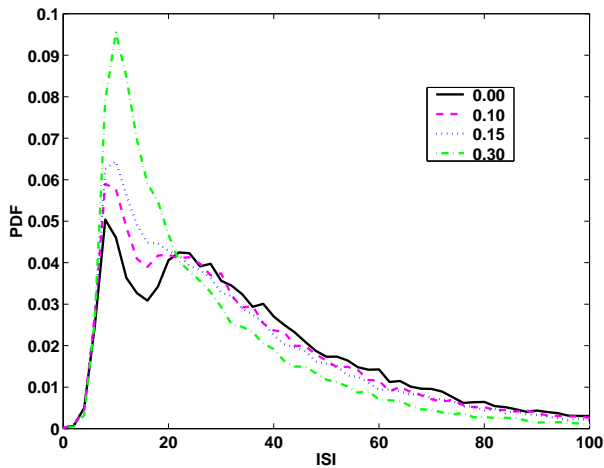}}
\caption{Probability distribution function of ISIs for $k=0.426$, $D=0.03$ and different values of $\tau$ (= 0.00, 0.10, 0.15 and 0.30) }
\label{pdf}
\end{figure}
The value of $D=0.03$ is chosen because it marks a region in parameter space where for $\tau=0$ spikes and bursts are nearly equally abundant. This is clearly seen in the solid curve of Fig. 3, which has a two-hump distribution with the first peak (low ISI) corresponding to bursting and the second one of nearly equal height corresponding to spiking (large ISI). Note that the second peak also has a long tail which is characteristic of spiking. With increasing values of $\tau$, the second peak gradually diminishes (the various dashed curves) and the long tail gets suppressed. On the other hand, the first peak becomes substantially stronger and for $\tau=0.3$, bursting behaviour clearly dominates.
\begin{figure}
\centerline{\includegraphics[width=8.0cm,angle=0]{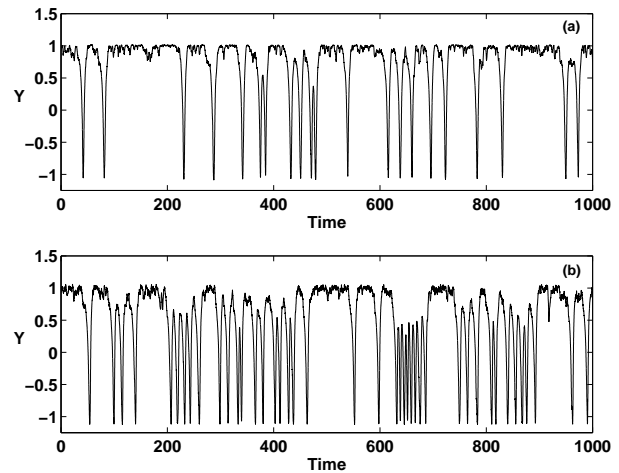}}
\caption{Typical time series for $k=0.426$ and $D=0.03$ for $(a)$ $\tau=0$ and $(b)$ $\tau=0.3$. Spikes dominate in $(a)$ whereas predominantly bursting events are seen in $(b)$}
\label{ts1}
\end{figure}
These results are also visually discernible in snapshots of segments of time series data corresponding to the $\tau$  values of $0$ and $0.3$ respectively in Fig.4a,b. \\

To understand the dynamical reason for this transition from spiking to bursting as a function of $\tau$, we have examined the effect of time delay on the characteristic properties of the limit cycle itself. We find that the primary influence is on the basic frequency of the limit cycle. A typical example is shown in Fig. 5 where the limit cycle frequencies for different values of the feedback strength $k$ are shown for $\tau=0$ (solid circles) and $\tau=0.3$ (solid line). We see that the frequencies are always higher at finite values of $\tau$. The physical consequence of this change in frequency is that the system now executes a larger number of cyclic orbits before returning to the quiescent state - which is characteristic of bursty behaviour. The consequent preponderance of small ISIs as compared to the no delay case is directly reflected in the probability distribution function curves shown in Fig. 3.  The sensitivity of the excitability property of the system to time delay is quite remarkable and is the other major interesting result of our present investigation. It suggests that  time delay can provide an effective mechanism for steering the system towards either a predominance of spiking or of bursting behaviour. Since the nature of excitability is at the heart of many important neuronal functions such as communication, computational properties and information processing our finding can have important practical consequences for the collective dynamics of such interacting neurons. As an example the process of noise coupled synchronization between trains of neuronal signals can be significantly influenced by changes in the intrinsic time delay parameter of each neuron and create interesting consequences for inter-neuron communication. Likewise the computational properties of a neuron can change due to its transition from a spiking to a bursting state brought about by the presence of time delay. \\

To conclude, in this paper we report two interesting results from the dynamical study of a noise driven model neuron that
consists of a subcritical Hopf oscillator with a time delayed nonlinear feedback. We find that the coherence resonance parameter of such a system possesses a bi-modal structure with the two peaks corresponding to optimal noise levels for two different kinds of excitations - namely single spikes and multi-spike (bursty) structures. Such a discriminatory ability enhances and to some extent generalizes the utility of the CR measure as a statistical tool for the analysis of neuronal data. Our other finding is that the presence of time delay in the nonlinear feedback can significantly influence the excitability properties of the system and bring about a controlled change in state from a predominantly spiking behaviour to largely bursting behaviour. This can have important practical consequences for various neuronal  mechanisms (e.g. communication, information processing, computation etc.) where action potentials come into play.\\
%\newpage 

\begin{figure}
\centerline{\includegraphics[width=8.0cm,angle=0]{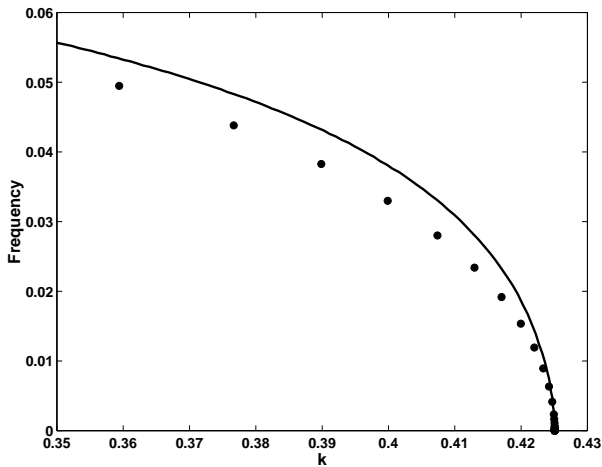}}
\caption{The frequencies of the limit cycles for $\tau=0$ (solid circles) and for $\tau=0.3$ (solid line). Note the increase in frequencies for a finite value of time delay.}
\label{ts2}
\end{figure}
%\paragraph{}
All the numerical simulations in the present paper have been carried out with the help of the software packages XPPAUT \cite{xppaut} and DDE-BIFTOOL \cite{dde}. XPPAUT also provides an interface to the continuation software package AUTO \cite{auto}. 
%\paragraph{}
The authors thank C.S. Zhou for valuable discussions. JK acknowledges the support via his Humboldt-CSIR research award.

%REFERENCE
%\bibliography{references}

\end{document}